\documentstyle[11pt,aasms4,flushrt]{article}

\newcommand{\apl}{\lesssim}
\newcommand{\cmjj}{\mbox{${\rm cm^{-2}}$}}
\newcommand{\etal}{et al.}
\newcommand{\hI}{\mbox{${\rm H\ I}$}}
\newcommand{\kms}{\mbox{km\ s${^{-1}}$}}
\newcommand{\lya}{\mbox{${\rm Ly}\alpha$}}
\newcommand{\lyb}{\mbox{${\rm Ly}\beta$}}
\newcommand{\civ}{\mbox{${\rm C\ IV}$}}
\newcommand{\cii}{\mbox{${\rm C\ II}$}}
\newcommand{\oi}{\mbox{${\rm O\ I}$}}
\newcommand{\ovi}{\mbox{${\rm O\ VI}$}}
\newcommand{\nii}{\mbox{${\rm N\ II}$}}
\newcommand{\nv}{\mbox{${\rm N\ V}$}}
\newcommand{\siiv}{\mbox{${\rm Si\ IV}$}}
\newcommand{\siii}{\mbox{${\rm Si\ II}$}}
\newcommand{\siiii}{\mbox{${\rm Si\ III}$}}

\def\N#1{{N({\rm #1})}}
\def\cm#1{\, {\rm cm^{#1}}}
\newcommand{\ibid}{\underline{\makebox[0.5in]{}}.} 

\begin{document}

\lefthead{Chen \& Prochaska}
\righthead{}

\slugcomment{Accepted by The Astrophysical Journal Letters}

\title{THE ORIGIN OF A CHEMICALLY ENRICHED \lya\ ABSORPTION SYSTEM AT 
$z=0.167$\altaffilmark{1}}
\altaffiltext{1}{Based on observations with the NASA/ESA Hubble Space
Telescope, obtained at the Space Telescope Science Institute, which is operated
by the Association of Universities for Research in Astronomy, Inc., under NASA
contract NAS5--26555.}

\author{HSIAO-WEN CHEN and JASON X. PROCHASKA}
\affil{Observatories of the Carnegie Institution of Washington, 813 Santa 
Barbara Street, Pasadena, CA 91101, U.S.A. \\
hchen,xavier@ociw.edu}

\newpage

\begin{abstract}

  We present the first detailed analysis of the chemical abundances, ionization
state, and origin of a partial Lyman limit system ($N(\hI) \approx 10^{16}$ 
\cmjj) at low redshift ($z=0.167$ towards PKS0405$-$1219).  Two galaxies at 
$\approx 70\ h^{-1}$ kpc projected distance to the QSO have been identified at 
the absorber redshift.  We analyze an echelle spectrum of the QSO obtained with
the Space Telescope Imaging Spectrograph and find that this absorption system 
exhibits associated lines produced by C$^+$, N$^+$, O$^0$, Si$^+$, Si$^{++}$, 
Si$^{+3}$, Fe$^+$, and Fe$^{++}$, and most interestingly, O$^{+5}$ and 
N$^{+4}$.  The results of our analysis show that the partial Lyman limit system
traced by various ions is likely to be embedded in a collisionally ionized 
\ovi\ gas of larger spatial extent.  Furthermore, the partial Lyman limit 
system appears to have a metallicity of {\em at least} $1/10$ solar and most 
likely solar or super solar despite the fact that no luminous galaxies are 
seen within a projected distance $\rho = 60\ h^{-1}$ kpc to the QSO.  Finally,
adopting the temperature estimated for the hot gas $T\approx 2.5\times 10^5$ K
and assuming a simple isothermal halo, we estimate that the galaxy or galaxy 
group that supports the extended gas may have a total mass $\approx 1.5\times 
10^{12} M_\odot$ and a gas number density $\apl 3\times 10^{-5}$ cm$^{-3}$.

\end{abstract}


\newpage

\renewcommand{\thefootnote}{\fnsymbol{footnote}}

\section{INTRODUCTION}

  The absorption line systems observed in the spectra of background QSOs have 
proven to be a sensitive probe to the physical conditions of intervening gas.  
In particular, kinematic studies and chemical abundance measurements of \lya\ 
absorption systems of neutral hydrogen column density $N(\hI)>10^{16}$ \cmjj\ 
at $z>1.7$ provide a direct assessment of the dynamical characteristics and
chemical enrichment history of the absorbers (Prochaska \& Wolfe 1997, 1998, 
1999, 2000; Pettini \etal\ 1997; Rauch, Haehnelt, \& Steinmetz 1997; Haehnelt, 
Steinmetz, \& Rauch 1998; Prochaska 1999).  While these absorption systems are 
generally believed to originate in or near galaxies (because of high \hI\ 
column density and/or large metal content), it has been extremely difficult to 
directly associate the properties of absorption line systems with the 
properties of galaxies because distant galaxies are faint.  On the other hand, 
at redshift $z\apl 1$, where galaxies are routinely identified, various galaxy 
surveys targeted at QSO fields have shown that luminous galaxies possess 
extended Mg II gas out to $\approx 40\ h^{-1}$ kpc\footnote{We adopt a standard
Friedmann cosmology of dimensionless Hubble constant $h = H_0/(100$ km s$^{-1}$
Mpc$^{-1})$ and deceleration parameter $q_0 = 0.5$ throughout this paper.} 
(e.g.\ Bergeron \& Boiss\'e 1991), extended \civ\ gas out to 100 
$h^{-1}$ kpc (Chen \etal\ 2000a), and extended neutral hydrogen gas out to 
$\approx 180\ h^{-1} $ kpc (Lanzetta \etal\ 1995; Chen \etal\ 1998, 2000b).  
But the origin and physical environment of the extended gas is largely unknown,
because of limited information regarding kinematics and metallicity of the gas.
The primary difficulty arises in acquiring a high-quality QSO spectrum of wide 
UV spectral coverage.

  To address this issue, we are initiating a survey of \lya\ absorption systems
for which high resolution, high signal-to-noise ratio (SNR) UV spectra are
available and for which the absorbing galaxies have been identified.  In this
{\em Letter}, we present the first results of this survey based on the study of
a \lya\ absorption system at $z = 0.167$ toward PKS0405$-$1219 ($z_{em} = 
0.5726$), for which high-quality echelle spectra have been obtained with the 
Hubble Space Telescope (HST) using the Space Telescope Imaging Spectrograph 
(STIS).  This system is especially interesting for two reasons.  First, 
previous measurements show that this system has a rest-frame \lya\ absorption 
equivalent width of 0.65 \AA, implying an \hI\ column density $N(\hI)=10^{15.3}
-10^{17.6}$ cm$^{-2}$ for a reasonable range of Doppler parameter (Chen 
\etal\ 2000b).  This suggests that the absorber is a partial Lyman limit system
at $z\approx 0$, which may provide a benchmark for comparison against Lyman 
limit systems (LLS) at high redshift.  Second, two galaxies have been 
identified at the absorber redshift (Spinrad \etal\ 1993; Chen \etal\ 2000b).  
A detailed study of this absorption system therefore serves to constrain the 
physical properties of the galaxy environment, which offers a unique 
opportunity to investigate whether QSO absorption line systems arise in 
extended halos of individual galaxies or in intragroup media of galaxy groups.

\section{ANALYSIS}

  Spectroscopic observations of PKS 0405$-$1219 were accessed from the HST data
archive.  The QSO was observed with HST using STIS in echelle mode with a 
$0''.2 \times 0''.06$ slit and the E140M grating ($R=45800$ or $6.7$ \kms) for 
a total exposure time of 27,208 s.  The observations were carried out in two 
visits of five exposures each.  The individual echelle spectra were reduced, 
extracted, and calibrated using standard pipeline techniques, and were coadded 
to form an averaged spectrum and a 1 $\sigma$ error array per visit using our 
own reduction program.  To form a final spectrum for absorption line studies, 
we normalized each coadded spectrum with a best-fit, low-order polynomial
continuum and calculated a weighted average of the normalized spectra with the 
weighting factor determined by the squares of the SNR.  The final spectrum 
covers a spectral range that spans from $\approx$ 1140 \AA\ to $\approx$ 1730 
\AA\ and has SNR of $\approx 7$ per resolution element in most of the spectral 
region.

  We identified absorption features produced by H$^0$, C$^+$, N$^+$, O$^0$, 
Si$^+$, Si$^{++}$, Si$^{+3}$, possibly Fe$^+$ and Fe$^{++}$; and most 
interestingly we identified the absorption doublets produced by N$^{+4}$ and 
O$^{+5}$ for the absorption system at $z=0.167$.  Figure 1 shows the velocity 
profiles of all transitions with $v = 0$ corresponding to redshift $z=0.1671$. 

  We first determined the \hI\ column density by fitting Voigt profiles to the
saturated \lya\ and \lyb\ absorption lines using the VPFIT package provided by 
R. Carswell and J. Webb.  But because of the degeneracy between $N(\hI)$ and 
$b$ for saturated Voigt profiles, we found that while the profile fitting 
procedure yielded a best fit\footnote{The solutions included additional 
components in the blue wing of the \lya\ line which do not impact the $N(\hI)$ 
of the absorption system at $z=0.167$.} $N(\hI) = 10^{15.8 \pm 0.2}$ \cmjj\ 
with $b = 35 $ \kms, solutions with similar reduced $\chi^2$ range from $\log 
N(\hI) = 15.7$ to $\log N(\hI) = 17.0$ for $b$ from 36 \kms\ to 26 
\kms.  Additional spectroscopic observations of the QSO in the far UV 
wavelength range (e.g.\ with the Far Ultraviolet Spectroscopic Explorer) are 
required to precisely measure the \hI\ column density using higher-order Lyman 
series absorption.  Despite the large uncertainty in the \hI\ column density, 
we confirmed that the absorber is a partial LLS with $N(\hI)\approx 10^{16}$ 
cm$^{-2}$.  Next, we measured the ionic column densities using the apparent 
optical depth method (Savage \& Sembach 1991), which provides an accurate 
column density estimate for resolved, unsaturated absorption lines.  The column
density measurements are presented in Table 1, which lists the ions, the 
corresponding rest-frame absorption wavelength $\lambda_0$, and the estimated 
ionic column densities $\log N$ together with the associated errors in the 
first three columns.  Lower limits indicate that the lines are saturated in the
STIS spectrum.

  The detections of N$^{+4}$ and O$^{+5}$ at the absorber redshift are very 
interesting, because these ions are often believed to form through collisional 
ionization in hot gas clouds.  We show in Figure 1 that all the ions except 
N$^{+4}$ and O$^{+5}$ have consistent profile signatures that trace the partial
LLS, while the \nv\ and \ovi\ doublets appear to be broad and have velocity 
centroids blue-shifted by $\approx$ 30 \kms\ from the other ions.  The 
differences in kinematic signatures strongly indicate that the \nv\ and \ovi\ 
doublets and the partial LLS traced by the other ions do not arise in the same 
regions.  For this reason, we studied the physical environment separately for 
the two gas regions.

  We first determined the temperature of the partial LLS using the $b$ 
parameters estimated from a Voigt profile analysis for the unsaturated lines 
produced by Si$^+$ and N$^+$.  The estimates of the $b$ parameters are
presented in the forth column of Table 1.  We found that $b=9.9 \pm 0.9$ \kms\ 
for Si$^+$ and $b=11.6 \pm 1.0$ \kms\ for N$^+$.  Because both thermal motion 
and bulk motion contribute to the measured Doppler parameter and because Si$^+$
and N$^+$ share the same bulk motion, we solved for $b_{\rm bulk}$ and $T$.  We
found that $b_{\rm bulk} = 6.8\pm 2.9$ \kms\ and $T\approx 7.4 \times 10^4$ K
with a 1 $\sigma$ lower limit being $T\approx 2.7 \times 10^4$ K and 1 $\sigma$
upper limit being $T\approx 1.5 \times 10^5$ K.  

  According to Sutherland \& Dopita (1993), collisional ionization models 
cannot produce the observed relative abundances for the Si ions at temperatures
between $T\approx 2.7 \times 10^4$ K and $T\approx 1.5 \times 10^5$ K.  It is
therefore very unlikely that the partial LLS is collisionally ionized.  We 
determined the ionization state of the partial LLS by comparing the column 
densities of H$^0$, Si$^{+}$, Si$^{++}$, and Si$^{+3}$ with the predictions 
from a series of photoionization models calculated using the CLOUDY software 
package (Ferland 1995).  Considering a plane-parallel geometry for gas of solar
metallicity, we calculated the column density predictions for various ions.  
The predictions of relative ionic column densities are fairly insensitive to 
the adopted metallicity and \hI\ column density (for optically thin gas) in the
CLOUDY calculations, even though the absolute predictions may vary accordingly.
Therefore, we were able to place reasonably tight constraints on the ionization
parameter, which is the ratio of incident ionizing photons to the total 
hydrogen number density, $U\equiv\phi_{912}/\,cn_{\rm H}$, by comparing the 
relative abundances of the Si ions.  We found $\log\,U = -2.64 \pm 
0.07$\footnote{The error in $U$ is based solely on the error in 
$\N{SiII}/\N{SiIV}$ and does not account for the systematic errors associated 
with the simplified assumptions inherent to the CLOUDY calculations.} using the
Si$^+/\,$Si$^{+3}$ ratio, which is in a good accordance with the limits derived
from Si$^{++}/\,$Si$^+$ and Si$^{+}/\,$Si$^{++}$.  Given the best estimated 
$U$, we determined the total hydrogen column density $N({\rm H})$, ionization 
fraction $x$, hydrogen number density $n_{\rm H}$, and the spatial extent for 
the partial LLS.  The results are summarized in Table 2.

  Finally, we estimated the elemental abundances of the partial LLS using the 
ionization fraction correction calculated from the CLOUDY model for an 
ionization parameter $\log\,U=-2.64$.  The results are shown in the last column
of Table 1 for an \hI\ column density $\log\,N(\hI)=16.5$.  Adopting the
largest possible \hI\ column density $N(\hI)=10^{17}$ cm$^{-2}$, we found that 
the partial LLS would have $1/10$ solar abundance determined for
carbon and silicon and $1/5$ solar abundance for nitrogen.  Adopting the best 
fit $N(\hI)$ of $10^{15.7} \cm{-2}$, we derived a super-solar metallicity for 
the system.  In addition, we also estimated the chemical abundance using the 
\oi\ absorption line.  In regions where the resonant charge-exchange reaction 
between oxygen and hydrogen becomes dominant, the column density ratio 
$[\oi/\,\hI]$ provides a direct estimate of the elemental abundance of oxygen. 
Otherwise, the ratio $[\oi/\,\hI]$ provides a lower limit to the estimate of 
oxygen elemental abundance.  We derived a metallicity of at least $1/2$ solar 
for oxygen and contend that the absorbing gas has been heavily enriched in 
metals.

  The results obtained from the CLOUDY analysis demonstrated that a single 
photoionization model cannot explain the observed column densities of the 
O$^{+5}$ and N$^{+4}$ ions.  At the extreme upper limit $\log\,U=-2.2$, both 
\ovi\ and \nv\ are predicted to be at least 0.4 dex less abundant than \siiv. 
But our measurements indicate that \ovi\ and \nv\ are at least 0.4 dex more 
abundant than all the Si ions.  In agreement with our assessment of the 
kinematic characteristics, the CLOUDY predictions further demonstrated that 
O$^{+5}$ and N$^{+4}$ do not arise in the partial LLS.  Under the assumption of
collisional ionization, the temperature of O$^{+5}$ and N$^{+4}$ may be derived
by comparing their column density ratio with a series of collisional ionization
models.  According to Shapiro \& Moore (1976), an optically thin gas of 
solar relative abundance at thermal equilibrium with $N(\ovi)/\,N(\nv) = 6.9 
\pm 1.5$ has a temperature of $T = (2.6\pm 0.1) \times 10^5$ K\footnote{The
estimated temperature may differ by only $\approx$ 5\% if the relative 
abundance of oxygen to nitrogen is twice the solar.}.  At this temperature, we 
calculated the total hydrogen column density using the ionization fraction 
correction calculated by Sutherland \& Dopita (1993).  We also estimated the 
bulk motion of these ions by fitting Voigt profiles to the \ovi\ and \nv\ 
absorption doublets.  The results are summarized in Table 2.

\section{DISCUSSION}
 
  Our analysis of the STIS echelle spectrum revealed unusual properties of the 
\lya\ absorption system at $z=0.167$.  We found a chemically enriched, warm 
gas region giving rise to the partial LLS and a hot gas region giving rise to
O$^{+5}$ and N$^{+4}$ along the QSO line of sight.  Two galaxies have been 
identified in the field of PKS0405$-$1219 at the absorber redshift (Spinrad 
\etal\ 1993; Chen \etal\ 2000b): (1) an elliptical galaxy at $z=0.1667$ with an
angular distance $\theta=40.4''$ to the QSO (corresponding to an impact 
parameter $\rho = 74.9\ h^{-1}$ kpc) and a rest-frame $K$-band luminosity $L_K 
= 1.20\ L_{K_*}$ and (2) a spiral galaxy at $z=0.1670$ with $\theta=33.9''$ 
(corresponding to $\rho = 62.8 \ h^{-1}$ kpc) and $L_K = 0.02\ L_{K_*}$.  While
it is not uncommon to find tenuous gas at large galactic distance ($\rho \sim 
100 \ h^{-1}$ kpc) around nearby elliptical galaxies (e.g.\ Fabbiano, Kim, \& 
Trinchieri 1992) or groups of galaxies (e.g.\ Mulchaey \etal\ 1996), it is very
surprising to find metal-enriched, high-column density gas at this large 
distance.  Furthermore, our column density estimates for O$^{+5}$ and N$^{+4}$ 
are comparable to the measurements obtained for highly ionized gas in the 
Galactic halo, most of which is believed to arise near the galactic plane 
(Savage, Sembach, \& Lu 1997; Savage \etal\ 2000).  The spectral 
characteristics of the elliptical galaxy exhibit signs of recent star 
formation (Spinrad \etal\ 1993), therefore the physical process that initiated
the recent star formation might be responsible for transporting metals to 
large galactic distance.

  In light of the different kinematic characteristics between the partial LLS 
and the \ovi\ absorbing cloud, it seems likely that the partial LLS is embedded
within a larger volume of hot \ovi\ gas that is spread out in a galactic halo 
or an intragroup medium.  But it is very rare to detect hot gas at a 
temperature $T\approx 2 \times 10^5$ K, where the cooling function is the most 
effective (see e.g.\ Spitzer 1978).  In order for the the hot gas to remain at 
this temperature, we derive rough estimates of the underlying mass and gas 
number density assuming a simple isothermal halo.  Specifically, a more 
massive system would have a higher virial temperature and consequently raise 
the gas temperature through gravitational interaction.  A higher gas density 
would increase the cooling rate and consequently lower the gas temperature 
through collisional cooling processes.  We estimate the mass of the galaxy or
galaxy group that supports the extended hot gas by requiring a virial 
temperature $T\approx 2.5 \times 10^5$ K and find that a total mass of $\approx
1.5\times 10^{12} M_\odot$ for a half-mass radius $r_h \approx 0.35$ Mpc.  We 
estimate the number density by requiring a cooling time longer than the 
dynamical time, $t_{\rm dyn}\approx \pi R/\,\sqrt 2 \sigma$, and find that the 
number density of the absorbing gas should be $\apl 3\times 10^{-5}$ cm$^{-3}$ 
for a velocity dispersion $\sigma\approx 60$ \kms\ at $R\approx 100$ kpc 
(implying $t_{\rm dyn}\approx$ 3.5 Gyr).

  Although two galaxies have been identified at the absorber redshift, it is
very difficult to determine whether the absorbing gas originates in extended 
halos of individual galaxies or in an intragroup medium of an underlying galaxy
group without a more complete spectroscopic survey of galaxies in this field.  
We can however derive constraints on the properties of the galaxies either 
individually or collectively for both scenarios based on all the available 
measurements.

  Comparison of galaxies and \lya\ absorption systems along common lines of 
sight has shown that the gaseous extent $r$ of galaxies of all morphological 
types scales with galaxy $K$-band luminosity $L_K$ according to $r\propto 
L_K^{0.28\pm 0.08}$ (Chen \etal\ 2000b).  According to these authors, we expect
that the extended gas of the spiral galaxy would contribute an \hI\ column 
density of no more than $3.2 \times 10^{14}$ cm$^{-2}$ at $\rho = 62.8\ h^{-1}$
kpc, but that the extended gas of the elliptical galaxy would easily contribute
an \hI\ column density of $10^{17}$ cm$^{-2}$ at $\rho = 74.9\ h^{-1}$ kpc.  
By carefully examining deep images obtained both with HST using the Wide Field 
and Planetary Camera 2 with the F702W filter and with the IRTF 3 m telescope 
using the K' filter, we find that if there are unidentified dwarf galaxies at 
the absorber redshift and closer to the QSO line of sight, then they cannot be 
brighter than $0.02\ L_{K_*}$.  It is therefore more likely that the absorbing 
clouds should follow the halo motion of the bright elliptical galaxy if the 
absorption system arises in individual galactic halos.  Assuming that the 
absorption system is produced in the gaseous halo of the elliptical galaxy and 
adopting a King profile for the gas distribution around the elliptical galaxy, 
$n(r)\propto 1/[1+(r/r_c)^2]$, we derive a total gas mass of $M_{\rm gas}\apl\ 
10^{10} M_\odot$ within a radius $\rho<100\ h^{-1}$ kpc for a core radius 
between $r_c = 10$ and 50 kpc and for a unit filling factor.  The estimated 
gas mass within $100\ h^{-1}$ kpc is significantly lower than the dynamical 
mass $3.6\times 10^{11} M_\odot$ estimated for $M/L_K\approx 6.6$, implying a 
small gas fraction in the inner part of the elliptical galaxy.

  Mulchaey \etal\ (1996) suggested that an intragroup medium of temperature
just below the detection threshold of existing X-ray observations may reveal 
itself through the imprint of high-ionization absorption lines, such as \ovi\ 
and Ne VIII, in the spectra of background QSOs.  Recent observations also 
indicated that some \ovi\ absorption lines originate in an environment of 
excess galaxy counts (Tripp, Savage, \& Jenkins 2000).  Adopting the $\sigma-T$
correlation obtained for groups of galaxies in X-ray (Mulchaey \& Zabludoff 
1998) and extrapolating to lower temperature, we find a corresponding velocity 
dispersion of $\sigma\approx$ 60 \kms\ for $T\approx 2.5\times 10^5$ K, which
is very close to the estimated bulk motion for the \ovi\ absorbing gas.  Based 
on the estimated velocity dispersion, we derive a total mass of $M\approx 1.5
\times 10^{12} M_\odot$ for the galaxy group assuming a virialized isothermal 
halo and a half-mass radius $r_h\approx 0.35$ Mpc.  Therefore, we find that 
if the absorption system is produced in the intragroup medium, then the total 
mass of the underlying galaxy group is comparable to the Local Group (Courteau 
\& van den Bergh 1999).

  The absorber at $z=0.167$ exhibits striking resemblance to the LLS at $z =
0.79$ presented by Bergeron \etal\ (1994) both in physical properties of the 
absorbing gas and in the surrounding galaxy environment.  The absorber at 
$z=0.79$ was found to have a metallicity of a half solar and likely be embedded
in a hot \ovi\ gas of $T \approx 2\times 10^5$ K.  A group of galaxies have 
been identified at the absorber redshift, the closest of which is at 110 
$h^{-1}$ kpc to the QSO line of sight.  If the two systems are representative 
of low-redshift LLSs, then the relationship between LLSs and surrounding 
galaxies may provide important clues regarding the chemical enrichment history
and dynamics of gas in intragroup media.  On the other hand, various models 
have been proposed to explain extended \hI\ gas at large galactic distance, 
including satellite accretion (Wang 1993), tidal debris (Morris \& van den 
Bergh 1994), and two-phase galactic halos (Mo \& Miralda-Escud\'e 1996).  Of 
these models, accretion from a star-forming satellite galaxy is the most likely
to explain the observed high-metallicity gas at large distance to 
regular-looking galaxies at $z=0.167$.  

  Finally, comparison of the absorber at $z=0.167$ and known high-redshift LLSs
(K\"{o}hler \etal\ 1999; Lopez \etal\ 1999; Prochaska 1999; Prochaska \& Burles
1999; Rauch \etal\ 1999) shows that all but one (Rauch \etal) of these Lyman 
limit absorbers have similar estimates of the ionization parameter $U$.  If the
majority of LLSs continue to exhibit $\log U \approx -2.5$ at all epochs, then 
it clearly calls for the question whether these systems share similar physical 
properties such as gas temperature.  

\acknowledgments
 
  It is a pleasure to thank Jacqueline Bergeron, Alan Dressler, Ken Lanzetta, 
Pat McCarthy, John Mulchaey, Michael Rauch, Ben Weiner, and Ray Weymann for 
helpful discussions.  We acknowledge the STIS GTO team for obtaining the QSO 
spectra (program identification number 7576).  JXP acknowleges support from a 
Carnegie postdoctoral fellowship and would like to thank ESO for their 
hospitality.

\newpage

\begin{deluxetable}{p{0.5in}lrrr}
\tablecaption{Estimates of Column Densities, Doppler Parameters, and 
Abundances} 
\tablewidth{0pt}
\tablehead{\colhead{Species} & \colhead{$\lambda_0$} & 
\colhead{log $N$ (cm$^{-2}$)} &  \colhead{$b$ (\kms)} &  
\colhead{[X/H]\tablenotemark{a}}}
\startdata
 \hI\    \dotfill & 1025.72 & $>$ 15.7 & $34.7 \pm 2.5$ & ... \nl
                  & 1215.67 & $>$ 15.7 & ...  & ... \nl
 \cii\   \dotfill & 1036.79 & $>14.10$ & ...  & ... \nl
                  & 1334.53 & 14.27 $\pm$ 0.09 & ...  & $-0.33$ \nl
 \nii\   \dotfill & 1083.99 & $>14.25$ & $11.6 \pm 1.0$ &  $0.27$ \nl
 \nv\    \dotfill & 1238.82 & 13.84 $\pm$ 0.07 & $50.2 \pm 7.0$ & ... \nl
                  & 1242.80 & 13.91 $\pm$ 0.06 & ...  & ... \nl
 \oi\    \dotfill & 1302.17 & 13.68 $\pm$ 0.14 & ...  & $0.25$ \nl
 \ovi\   \dotfill & 1031.93 & 14.67 $\pm$ 0.16 & $72.6 \pm 4.9$ & ... \nl
                  & 1037.62 & 14.76 $\pm$ 0.07 & ...  & ... \nl
 \siii\  \dotfill & 1190.42 & 13.22 $\pm$ 0.07 & $9.9 \pm 0.9$  & ... \nl
                  & 1193.29 & 13.29 $\pm$ 0.05 & ...  & ... \nl
                  & 1260.42 & $>13.16$ & ...  & ... \nl
                  & 1304.37 & 13.40 $\pm$ 0.12 & ...  & $-0.37$ \nl
 \siiii\ \dotfill & 1206.50 & $>13.33$ & ...  & $>-0.76$  \nl
 \siiv\  \dotfill & 1393.76 & 13.18 $\pm$ 0.05 & ...  & ... \nl
                  & 1402.77 & 13.49 $\pm$ 0.07 & ...  & $-0.45$ 
\tablenotetext{a}{$[{\rm X}/\,{\rm H}]$ is defined as $\log\,[N({\rm X})/\,N(
{\rm H})] - \log\,[{\rm X}/\,{\rm H}]_\odot$.  We assumed $\log\,N(\hI)=16.5$ 
for the abundance estimaates.  Increasing (decreasing) $N(\hI)$ by 0.5 dex 
decreases (increases) the abundance measurements by 0.5 dex.}
\enddata
\end{deluxetable}

\newpage

\begin{deluxetable}{p{2.5in}rr}
\tablecaption{Physical Parameters of the Absorption System at $z=0.167$}
\tablewidth{0pt}
\tablehead{\colhead{Physical Parameter}&\colhead{Partial LLS\tablenotemark{a}} 
& \colhead{\nv, \ovi\tablenotemark{b}}}
\startdata
Ionization parameter $\log\,U$ \dotfill & -2.64 $\pm$ 0.07 & ... \nl
Ionization fraction $x$ \dotfill & 0.996 & ... \nl
Hydrogen column density $\log\,N({\rm H})$ \dotfill & 18.86 $\pm$ 0.07 & 
$\approx$ 18.6 \nl
Hydrogen number density $n_{\rm H}$ (cm$^{-3}$) \dotfill & $6\times 10^{-4}$ & 
$< 3\times 10^{-5}$  \nl
Cloud size $l$ (kpc) \dotfill & $\approx$ 4 & ... \nl
Bulk motion $b_{\rm bulk}$ (\kms) \dotfill & $\approx$ 7 & $\approx$
70 \nl
Temperature $T$ (K) \dotfill & $7.4\times 10^4$ & $2.5\times 10^5$ \nl
Chemical abundance $[{\rm Z}/\,{\rm H}]$ \dotfill & $>-0.1$ & ... 
\tablenotetext{a}{$N({\rm H})$,$n_{\rm H}$, and $l$ were estimated by adopting 
an ionizing radiation intensity $J_{912}\approx 2\times 10^{-23}$ ergs s$^{-1}$
cm$^{-2}$ Hz$^{-1}$ sr$^{-1}$ at $z=0$ estimated by Shull \etal\ (1999).}
\tablenotetext{b}{$N({\rm H})$ was estimated based on the measured \ovi\ 
column density, corrected for the ionization fraction of oxygen at $T\sim 5.4
\times 10^5$ K assuming a solar abundance. $n_{\rm H}$ was estimated by 
requiring a cooling time comparable to or greater than the dynamical time.}
\enddata
\end{deluxetable}

\newpage

\figcaption{Velocity profiles of absorption lines identified for the absorption
system at $z=0.167$ in the STIS spectrum with $v = 0$ corresponding to redshift
$z=0.1671$.  The identification of each transition is indicated in the lower
right corner of each panel.  The upper and bottom dashed lines indicate the 
continuum and zero levels, respectively.  The dash-dotted lines indicate the
velocity centroids.  Contaminating features from other sources that are 
excluded in our analysis are indicated by dotted curves. }

\newpage

\plotone{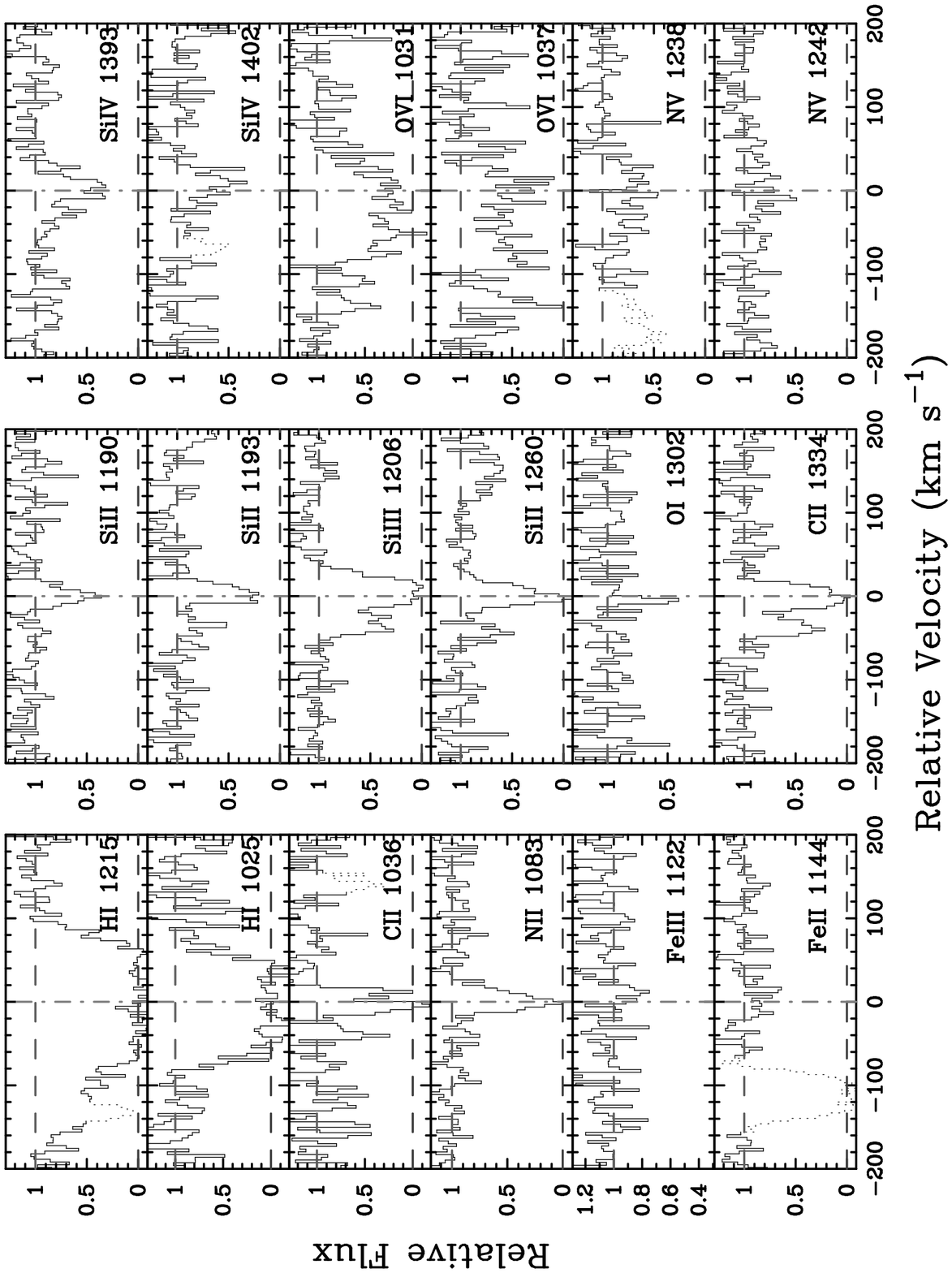}

\end{document}